\pdfoutput=1
\documentclass[twocolumn,showpacs,preprintnumbers,amsmath,amssymb]{revtex4}

\usepackage{graphicx}
\usepackage{dcolumn}
\usepackage{bm}

\begin{document}


\title{
Numerical study of the Kerr solution in rotating 
coordinates}

\author{S. Bai}
\author{G. Izquierdo}
 \altaffiliation[Current address: ]{Facultad de Ciencias, Universidad 
 Aut\'onoma del Estado de M\'exico, Toluca 5000, Instituto literario 
 100, Edo. Mex., M\'exico.}
\author{C. Klein}%
 \email{christian.klein@u-bourgogne.fr}
\affiliation{Institut de Math\'ematiques de Bourgogne,
		Universit\'e de Bourgogne, 9 avenue Alain Savary, 21078 Dijon
		Cedex, France}%

\date{\today}

\begin{abstract}
The Kerr solution in coordinates corotating with the horizon is 
studied as a testbed for a spacetime with a helical Killing vector in 
the Ernst picture. 
The solution is numerically constructed by solving the Ernst equation 
with a spectral method and a Newton iteration. We discuss convergence 
of the iteration for several initial iterates and different values of 
the Kerr parameters. 
\end{abstract}

\maketitle

\section{Introduction}\label{sec:intro}
Binary black holes in the latest stage before an eventual merger were 
generally seen as the most promising sources of gravitational waves 
to be detected with current ground based interferometers, and this 
has been done just recently in \cite{LIGO}. Solutions to the 
Einstein equations for such configurations can only be found 
numerically with current knowledge, and considerable progress has 
been made in the last decade in this context, see for instance \cite{sper} for a 
recent review of the field. It is generally assumed that there is a 
quasi-stationary phase of such a system before the inspiral where the 
change of the radius of the binary orbit due to emitted gravitational 
radiation is relatively small during one complete turn. Detweiler 
\cite{detweiler1,blackdet,detweiler2} suggested to approximate 
this quasi-circular phase by a system where the outgoing radiation is 
exactly compensated by incoming radiation. This approximation had 
been previously used for binary charges of opposite sign in Maxwell theory by 
Sch\"onberg 
\cite{schoenberg} and Schild \cite{schild}. 

In a general relativistic context, this approximation corresponds to 
the presence of a helical Killing vector. For the phase of 
quasicircular orbits of binary systems, this concept has proven very 
fruitful in numerical computations, see for instance 
\cite{friedman,ggb1,ggb2,andrade,CCGP,ULFGS,UF,BGLMW,UGM,YBRUF,BBP,BOP,BBHP} 
and references therein. Interesting numerical concepts have been 
developed in these references, see also \cite{LP1,LP2,BJN}. Helical 
Killing vectors have also proven to be useful in post-Newtonian 
calculations \cite{TBW,Tiec,GT}.

Spacetimes with a helical Killing vector are also interesting from a mathematical 
point of view. If such a Killing vector is global, the spacetime 
cannot have a regular null infinity, see \cite{gibstew,ashxan}. Loosely 
speaking the reason for this is that the incoming radiation needed to 
compensate the outgoing radiation in a nonlinear theory does not 
allow for a regular null infinity. As discussed for instance in 
\cite{kleinprd} in a formal expansion, invariants of the Weyl tensor 
should have an oscillation point at null infinity. A characteristic 
feature of a 
helical Killing vector is the change of sign of his norm at the 
\emph{light cylinder}, a  surface of cylindrical topology. This can be understood most easily in 
Minkowski spacetime where $\xi=\partial_{t}+\Omega\partial_{\phi}$ 
with $\Omega=const$ is 
a helical Killing vector in standard cylindrical coordinates, 
i.e., one just passes into a rotating frame. The norm of the Killing 
vector is in this case $f=1-\Omega^{2}\rho^{2}$, and the light 
cylinder corresponds to $\rho_{c}=1/\Omega$. 
In the rotating frame, the helically reduced flat d'Alembert operator 
reads
\begin{equation}
    \mathcal{L}:=\partial_{\rho\rho} +\frac{1}{\rho}\partial_{\rho}+\partial_{zz}
+(1-\Omega^{2}\rho^{2})\frac{1}{\rho^{2}}\partial_{\phi\phi}
    \label{eq:operator}.
\end{equation}
Inside the light cylinder the operator is elliptic, outside 
hyperbolic.  Such spacetimes with signature changes appear also in 
other relativistic contexts, see \cite{stewart}. Equations of mixed 
type are often a consequence of symmetry reductions as here where the 
norm of the Killing vector changes sign, see \cite{sonic}. Operators 
of the type (\ref{eq:operator}) belong to the symmetric positive 
equations discussed by \cite{friedrich} and \cite{lax}. Questions of 
existence and uniqueness of solutions for equations in the context 
of helical Killing vectors were discussed in \cite{torre,Torre2}, and 
in a general context in \cite{Otw,Otwbook,Otwbook2}. Concrete 
examples for helical Killing vectors in various settings were 
discussed in \cite{BHS,BS,BST}.

In \cite{kleinprd} the existence of a helical Killing vector in a 
spacetime was to used to factorize the metric with respect to the 
symmetry in a projection formalism first applied by Ehlers 
\cite{ehlers}, see also \cite{geroch,maisonehlers}. In this case the 
Einstein equations can be written in the form of a complex Ernst 
equation \cite{ernst} which replaces the constraint equations in a 
standard $3+1$ decomposition. The remaining Einstein equations 
describe a model of 3-dimensional gravity coupled to a sigma model, 
see \cite{sigma,ernstbook}. 

This rather elegant form of the equations has the disadvantage that 
the Killing horizons and the light cylinder are singularities of the 
equations. Thus it is not clear whether they are useful for numerical 
computations. To address this question, we study in this paper for a simple 
test case whether the numerical issues in this formalism can be 
surmounted. To this end we consider the exact Kerr solution in a 
frame corotating with the horizon. In this frame, the norm of the 
helical Killing vector vanishes at the horizon and at the light 
cylinder. The 3-metric is prescribed, thus the only equation to be 
solved is the Ernst equation.

To solve the latter, we use a finite computational domain between the 
horizon and an outer radius where the exact Kerr solution is 
imposed as boundary values. The equations are solved with a spectral 
method and a Newton iteration. The Komar integral is imposed in the 
iteration to address non-uniqueness issues. It is shown that the 
iteration converges rapidly unless the light cylinder is close to the 
computational boundary. In this case the iteration is amended with an 
Armijo scheme \cite{armijo}. 

The paper is organized is follows: In secction 2, we briefly review 
the projection formalism and the Ernst equation. In section 3 we give 
the Ernst potential for the Kerr solution in coordinates corotating 
with the horizon. In the same coordinates, the Ernst equation is 
formulated in section 4. The used numerical approaches for the paper 
are presented in section 5. In section 6 we discuss the convergence 
of the scheme for various initial iterates and parameters of the Kerr 
solution. We add some concluding remarks in section 7.

\section{Quotient space metrics and Ernst equations}
In this section we briefly summarize the approach to binary black 
hole spacetimes with a helical Killing vector of \cite{kleinprd} 
based on quotient space metrics first used in \cite{ehlers} (see 
also \cite{geroch}) in the form of  \cite{maisonehlers} and Ernst equations. 
The existence of a Killing vector $\xi$, in adapted coordinates  
$\xi=\partial_{t}$ where $t$ is not necessarily a timelike 
coordinate, can be used to establish a simplified 
version of the field equations by dividing out the group action. 
The norm of the Killing vector will be denoted by $f$. 

In this approach, the metric is written in the form 
\begin{equation}
    ds^{2}=-f(dt+k_{a}dx^{a})(dt+k_{b}dx^{b})+\frac{1}{f}h_{ab}dx^{a}
    dx^{b}
    \label{maison1};
\end{equation}
latin indices always take the values $1,2,3$ corresponding to the 
spatial coordinate.  Note that this decomposition is not defined at the fixed points of the group 
action, i.e.\ the zeros of $f$. The Einstein equations will be singular 
at the set of zeros of  $f$ which is also a problem for a numerical 
treatment. Note that this will not be the case in a standard $3+1$ 
decomposition of spacetime. But as will be shown in this paper,  the 
related numerical issues can be controlled, and thus the simplicity 
of the quotient space approach with respect to a standard $3+1$ 
decomposition can be also used in a numerical approach. 

The Einstein equations in vacuum can be put into the form of the 
complex Ernst equation
\begin{equation}
    fD_{a}D^{a}\mathcal{E}=D_{a}\mathcal{E}D^{a}\mathcal{E}
    \label{ein4}.
\end{equation}
Here 
the complex Ernst potential is given by $\mathcal{E}=f+ib$ 
\cite{ernst}, where $D_{a}$ denotes the covariant 
derivative with respect to $h_{ab}$, where the twist potential 
$b$ is defined via ($\epsilon^{abc}$ is the tensor density with 
$\epsilon^{123}=1/\sqrt{h}$)
\begin{equation}
    k^{ab}=\frac{1}{f^{2}}\epsilon^{abc}b_{,c}
    \label{maison4},
\end{equation}
where $h$ is the 
determinant of $h_{ab}$, where $k_{ab}=k_{b,a}-k_{a,b}$, and where 
all indices are raised and lowered with $h_{ab}$.

The equations for the metric $h_{ab}$ can be written in the form
\begin{equation}
    R_{ab}=\frac{1}{2f^{2}}
    \Re (\mathcal{E}_{,a}\bar{\mathcal{E}}_{,b})
    \label{ein1a},
\end{equation}
where $R_{ab}$ is the three-dimensional Ricci tensor corresponding to 
$h_{ab}$. 
It is obvious that zeros of the norm of the 
Killing vector are singular points of the equations. 

Thus the equations for the metric function $h_{ab}$ are the three 
dimensional Einstein equations with some energy momentum tensor which 
is a so-called sigma model, see the discussion in \cite{kleinprd} 
and references therein. Thus one can introduce a $2+1$ decomposition 
of the quotient space, preferably a foliation with respect to some 
coordinate $r$ in which the horizons of the black holes are constant 
$r$ surfaces. It is well known that the 6 equations (\ref{ein1a}) 
split in this case into 3 `evolution equations' containing second 
order derivatives with respect to this coordinate $r$, and 3 
`constraints' containing at most first derivatives with respect to 
$r$, see \cite{kleinprd} for the helical case. If both of these pairs of equations are satisfied on the 
horizons, it will be sufficient to solve one of them in 
the space in between. 

Thus the problem for binary black holes with a 
helical Killing vector is reduced to solving the Ernst equation 
(\ref{ein4}) and 3 of the equations (\ref{ein1a}) which are of first 
order in the derivatives with respect to this coordinate $r$ (which 
need not be related to spherical coordinates). All these equations 
are singular at the zeros of the norm $f$ of the Killing vector, the 
horizons and the light cylinder. In order to solve these 
equations numerically, one is faced with a singular boundary value 
problem with a singular light cylinder the position and form of which 
is not known a priori. In particular it has cylindrical topology and 
can thus not be an $r=const $ surface. 

As a test problem for these issues which is analytically known, 
we will study in this paper the Kerr black hole in a frame corotating 
with the horizon. The Kerr spacetime is stationary and axisymmetric, 
and both Killing vectors $\partial_{t}$ and $\partial_{\phi}$ in an
asymptotically non rotating coordinate system are commuting. This 
means that $\xi=\partial_{t}+\Omega\partial_{\phi}$ is also a Killing 
vector for arbitrary constant value of $\Omega$. We will consider the value of 
$\Omega$ for which the norm of $\xi$ vanishes at the horizon of the 
Kerr black hole. In the stationary axisymmetric case, the metric 
$h_{ab}$ can be chosen to be diagonal with a single unknown function, see for instance 
\cite{exac,ernstbook}.  This function can be obtained via a 
line integration in closed form. With this function given, the task is reduced 
to solve the Ernst equation. The vector $k_{a}$ can be chosen to have 
just a $\phi$ compoment which will be denoted by $a$. 
The model has a vanishing $f$ at the horizon for the 
Killing vector $\partial_{t}$, a light cylinder and the 
asymptotically rotating coordinate system. Thus a numerical approach 
to reproduce the Kerr solution in this setting could be 
possibly extended to the case of binary black hole spacetimes with a 
helical Killing vector. We treat this problem on a finite 
computational domain bounded on one side by the black hole horizon 
and on the other by the exact Kerr solution which we impose for 
simplicity as a boundary condition. In the general case one would 
impose instead boundary values 
inferred from the asymptotic  behavior of the solution, for instance 
a solution to the linearized Einstein equations as discussed in 
\cite{kleinprd}. 

Note that we only consider the Ernst equation here since it already 
has all relevant features we want to test for the binary case with a 
helical Killing vector. The metric $\mathrm{h}$ is given for a known 
Ernst potential via a first order equation. 
In the iterative approach to solve the equations we 
are applying here, this means that in each step of the iteration the 
metric $\mathrm{h}$ will be obtained via a quadrature. Thus to test 
the approach for the Kerr solution, it is sufficient to give the 
exact metric $\mathrm{h}$ and to solve merely the Ernst equation.

\section{Kerr solution in rotating coordinates}
In this section we give the Kerr solution in the Ernst formalism in 
coordinates corotating with the horizon. In \emph{Boyer-Lindquist 
coordinates} $r,\theta,\phi$,  the Kerr solution for a 
single black hole with mass $m$ and angular momentum 
$J=m^{2}\sin\varphi$ takes the form (see for instance \cite{exac} 
and references therein)
\begin{eqnarray}
    f&=&\frac{r^{2}-2mr+m^{2}\sin^{2}\varphi\cos^{2}\vartheta}{
    r^{2}+m^{2}\sin^{2}\varphi\cos^{2}\vartheta}\nonumber,\\ 
    b&=&-\frac{2m^{2}\sin\varphi\cos\vartheta}{r^{2}+m^{2}\sin^{2}\varphi\cos^{2}\vartheta}
    \label{boyer3a}
\end{eqnarray}
and
\begin{equation}
    a=\frac{2m^{2}\sin\varphi r \sin^{2}\vartheta}{
    r^{2}-2mr+m^{2}\sin^{2}\varphi\cos^{2}\vartheta}
    \label{boyer3b}.
\end{equation}
 The parameter $\varphi$ varies 
between 0, the Schwarzschild solution, and $\pi/2$, the extreme Kerr 
solution. 
The metric $\mathrm{h}$ reads
\begin{eqnarray}
    h_{rr} & = & =\frac{r^{2}-2mr+m^{2}
    \sin^{2}\varphi\cos^{2}\vartheta}{(r-R)\left(r-2m\sin^{2}\frac{\varphi}{2}\right)},
    \nonumber  \\
    h_{\vartheta\vartheta} & = & r^{2}-2mr+m^{2}
    \sin^{2}\varphi\cos^{2}\vartheta,
    \nonumber  \\
    h_{\phi\phi} & = & (r-R)\left(r-2m\sin^{2}\frac{\varphi}{2}\right)\sin^{2}\vartheta
    \label{boyer3}.
\end{eqnarray}

The horizon is located in these coordinates at 
$R=2m\cos^{2}\frac{\varphi}{2}$. At the horizon we have for 
(\ref{boyer3b}) 
\begin{equation}
    a=-1/\Omega_{BH}=-2m\cot \frac{\varphi}{2}
    \label{OmegaBH}
\end{equation}
where $\Omega_{BH}$ is the 
angular velocity that can be attributed to the horizon with respect 
to an  observer at infinity. The metric function $h_{\phi\phi}$ vanishes 
at the horizon. 

The solution is here given in an asymptotically non rotating frame. 
Using a transformation of the form $\phi'=\phi+\Omega t$, we 
get 
\begin{eqnarray}
    g_{00}' & = & g_{00}+2\Omega g_{03}+\Omega^{2}g_{33},
    \nonumber  \\
    g_{03}' & = & g_{03}+\Omega g_{33}.
    \label{kerr2}
\end{eqnarray}
For $f$ and $a$ this implies
\begin{equation}
    f'=f(1+\Omega a)^{2}-\Omega^{2} h_{\phi\phi}/f,\quad
    a'f'=af(1+\Omega a)-\Omega h_{\phi\phi}/f.
    \label{quasi3a}
\end{equation}
In 
corotating coordinates ($\Omega=\Omega_{BH}$), we have for 
(\ref{boyer3b})
\begin{equation}
    1+\Omega a= \frac{(r-R)(r-2m\sin^{2}\frac{\varphi}{2}
    \cos^{2}\vartheta)}{r^{2}-2mr+m^{2}\sin^{2}\varphi\cos^{2}\vartheta}
    \label{boyer3b1}.
\end{equation}
Thus $f'$
vanishes in coordinates corotating with the horizon as $r-R$ for 
$r\to R$
because of the linear term in $h_{\phi\phi}$. The term $(1+\Omega 
a)^{2}$ is quadratic in $r-R$.

For the Kerr solution in Boyer-Lindquist coordinates, eq. 
(\ref{quasi3a}) implies with $\tilde{r}=r/R$
\begin{align}
    f' & = 
    \frac{\tilde{r}-1}{\tilde{r}^{2}+\tan^{2}\frac{\varphi}{2}\cos^{2}\vartheta}\left\{-\sin^{2}\frac{\varphi}{2}\cos^{2}\frac{\varphi}{2}\sin^{2}\vartheta (\tilde{r}^{3}+\tilde{r}^{2})\right.
    \nonumber\\
    &+\left(1-\sin^{2}\frac{\varphi}{2}\sin^{2}\vartheta-\sin^{4}\frac{\varphi}{2} \sin^{2}\vartheta \cos^{2}\vartheta\right) \tilde{r}
         \nonumber\\
 & 
     \left.     -\tan^{2}\frac{\varphi}{2}\cos^{2}\vartheta+\tan^{2}\frac{\varphi}{2}\sin^{4}\frac{\varphi}{2}\sin^{2}\vartheta\cos^{2}\vartheta\right\}
    \label{fp}
\end{align}
and
\begin{align}
    a'f' & = 
    \frac{2m\tan\frac{\varphi}{2}\sin^{2}\vartheta(\tilde{r}-1)}{\tilde{r}^{2}+\tan^{2}\frac{\varphi}{2}\cos^{2}\vartheta}\left\{-\cos^{4}\frac{\varphi}{2}(\tilde{r}^{3}+\tilde{r}^{2})\right.
    \nonumber\\
     & 
     \left.-\cos^{2}\frac{\varphi}{2}\left(1+\sin^{2}\frac{\varphi}{2}\cos^{2}\vartheta\right) \tilde{r}
     +\sin^{4}\frac{\varphi}{2}\cos^{2}\vartheta\right\}
    \label{afp}
\end{align}
The function $f'$ in (\ref{afp}) is shown for $\varphi=1$ in 
Fig.~\ref{f_{1}}.
\begin{figure}[htb!]
   \includegraphics[width=0.49\textwidth]{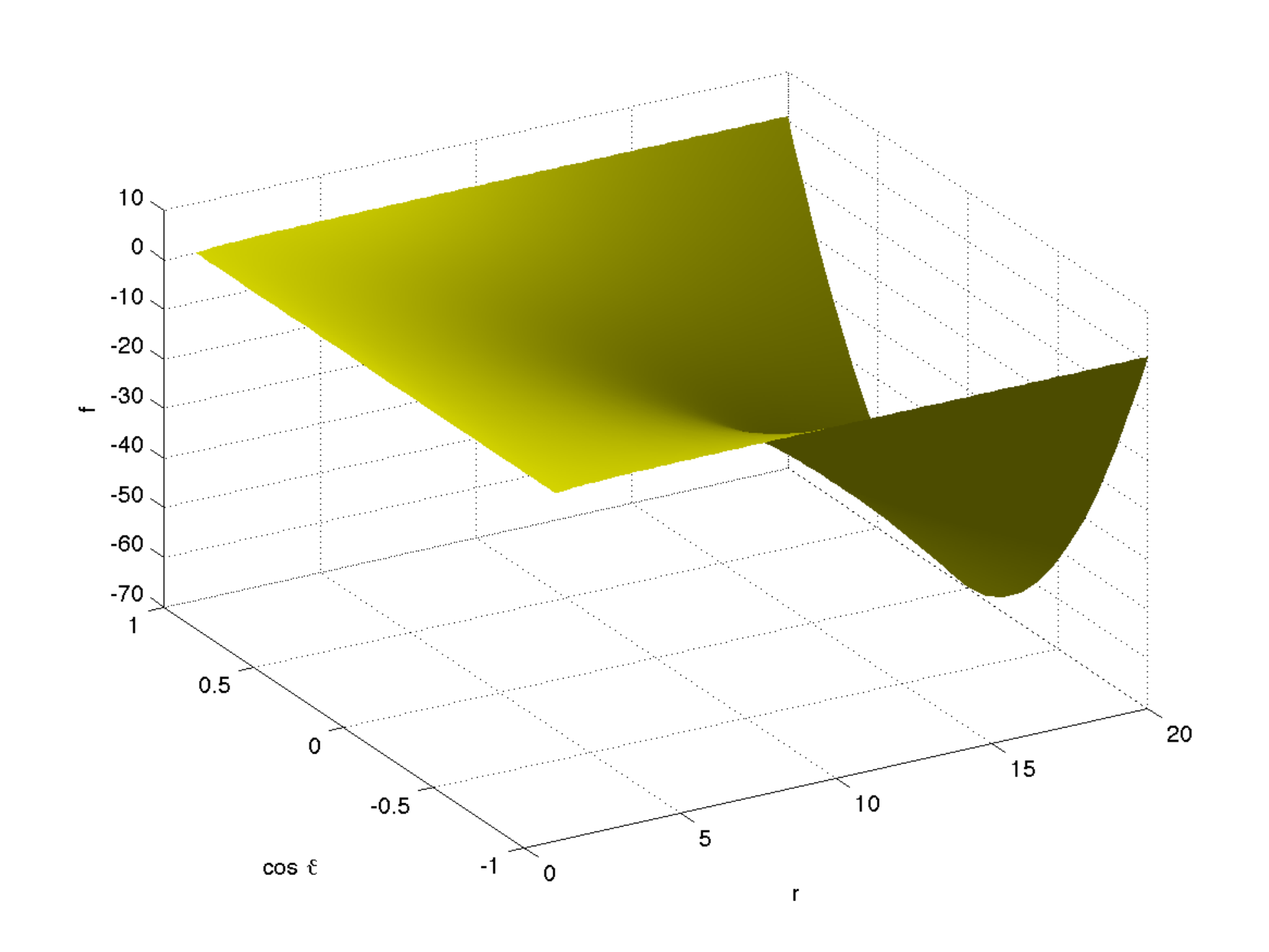}
 \caption{Real part of the Ernst potential for the Kerr solution in 
 coordinates corotating with the horizon (\ref{fp}) for $\varphi=1$. }
 \label{f_{1}}
\end{figure}

For the derivatives of \( b \) in (\ref{maison4}) one finds in 
Boyer-Lindquist coordinates 
\begin{equation}
    b'_{r}=\frac{a'_{\vartheta}f'^{2}}{((r-m)^{2}-m^{2}\cos^{2}\varphi)\sin\vartheta}
    \label{br}
\end{equation}
and
\begin{equation}
    b'_{\vartheta}=- \frac{a'_{r}f'^{2}}{\sin\vartheta}
    \label{btheta}.
\end{equation}
Here and in the following the index in $b'_{r}$ denotes the partial 
derivative with respect to the coordinate, here $r$. 
Integrating we find
\begin{equation}
    b' = 
    \frac{\sin\frac{\varphi}{2}\cos\frac{\varphi}{2}\cos\vartheta(\tilde{r}-1)^{2}\left(\tan^{2}\frac{\varphi}{2}\cos^{2}\vartheta-2\tilde{r}-1\right)}{
    \tilde{r}^{2}+\tan^{2}\frac{\varphi}{2}\cos^{2}\vartheta}
    \label{bp}.
\end{equation}
Obviously an integration constant was chosen such that $b'$ has a zero of second order at the 
horizon. For $r\to \infty$, it is proportional to $r\cos\vartheta$. 
It is a smooth function for all $\tilde{r}>1$.
The function is shown for $\varphi=1$ in Fig.~\ref{b_{1}}.
\begin{figure}[htb!]
   \includegraphics[width=0.49\textwidth]{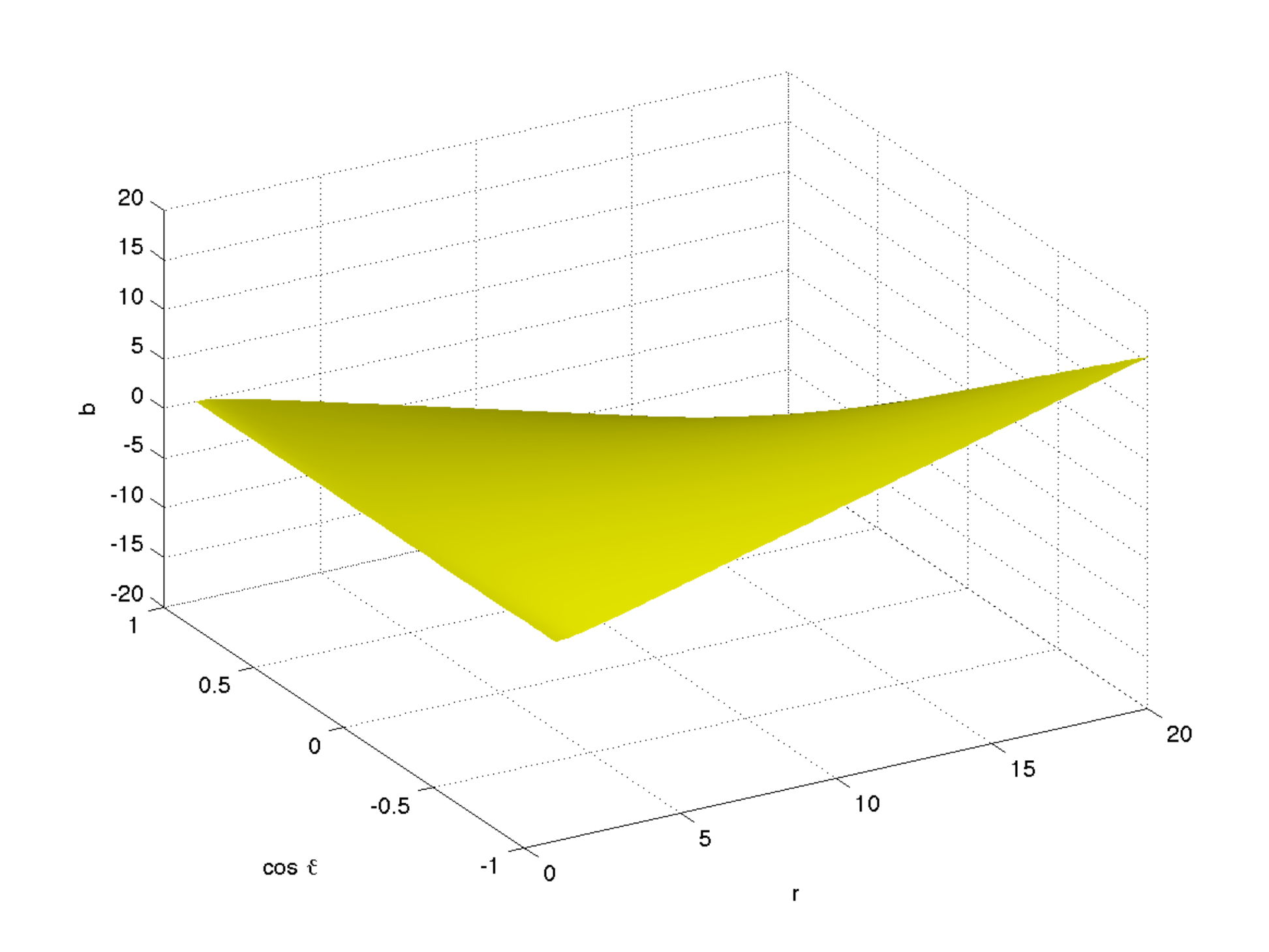}
 \caption{Imaginary part of the Ernst potential for the Kerr solution in 
 coordinates corotating with the horizon (\ref{bp}) for $\varphi=1$. }
 \label{b_{1}}
\end{figure}

The remaining metric functions of $\mathrm{h}$ are changed by 
multiplication with a factor $f'/f$. It is obvious that $f'=a'f'=0$ on 
the horizon. For large values of $r\sin \theta$, the norm $f'$ of the Killing 
vector becomes negative at  the light cylinder. All non-vanishing components 
of $\mathrm{h}$ except $h_{\phi\phi}'=h_{\phi\phi}$ will vanish 
there as $f'$. 
The function $b'$ is as expected odd in $\Omega$ and $\zeta$ and 
grows linearly in $\zeta$ for $r\to\infty$. The functions $a'f'$ and 
$f'$ grow as $\rho^{2}$. These kinematic contributions due to the asymptotically 
rotating coordinate system will be also present in the case with a 
helical Killing vector. For completeness we note the asymptotic behavior 
of the metric functions for the Kerr solution 
(which also holds for general asymptotically 
flat spacetimes) in Boyer-Lindquist coordinates,
\begin{eqnarray}
    f & = & 1-\frac{2m}{r}+0(1/r^{3}),
    \nonumber  \\
    b & = & -2m^{2}\sin\phi\frac{\cos\vartheta}{r^{2}}(1+0(1/r^{2})),
    \nonumber  \\
    a & = & 2m^{2}\sin\phi\frac{\sin^{2}\vartheta}{r}(1+0(1/r)),
    \label{asym1}.
\end{eqnarray}
\begin{widetext}
This implies in corotating coordinates
\begin{equation}
    f'  = 
    \left(1-\frac{1}{\tilde{r}}-\frac{\tan^{2}\frac{\varphi}{2}\cos^{2}\vartheta}{\tilde{r}^{2}}\right)\left\{-\sin^{2}\frac{\varphi}{2}\cos^{2}\frac{\varphi}{2}\sin^{2}\vartheta (\tilde{r}^{2}+\tilde{r})
    +1-\sin^{2}\frac{\varphi}{2}\sin^{2}\vartheta-\sin^{4}\frac{\varphi}{2} \sin^{2}\vartheta \cos^{2}\vartheta +0(1/\tilde{r})\right\}
    \label{asym2},
\end{equation}
and 
\begin{equation}
    b'=
    -2\sin\frac{\varphi}{2}\cos\frac{\varphi}{2}\tilde{r}\cos\vartheta\left(1-\frac{1}{2\tilde{r}}(3+\tan^{2}\frac{\varphi}{2}\cos^{2}\vartheta)+0(1/\tilde{r}^{2}))\right)
    \label{asym2b},
\end{equation}
up to some irrelevant constant. Note that $m^{2}\sin \phi$ is just the 
angular momentum. If one replaces this quantity by $J$, one gets the 
behavior in the general case.
\end{widetext}

Note that the Ernst potential for the Kerr solution is often given in 
so called Weyl coordinates since it takes a particularly simple form in these 
coordinates. But in Weyl coordinates, the metric functions are not 
regular at the horizon (they have a cusp like behavior where the axis 
intersects the horizon). Since the spectral methods we apply in this 
paper for the numerical solution of the Ernst equation are best 
adapted to analytic functions, we do not use this form of the 
coordinates here.

\section{Ernst equation in corotating coordinates and Komar integral}
In this section we formulate the Ernst equation in Boyer-Lindquist 
coordinates corotating with the horizon. This equation will be solved 
numerically in the following sections.  At some outer radius we impose 
the exact Kerr solution in rotating coordinates as a boundary 
condition. At the horizon, the vanishing 
of the norm of the Killing vector will be imposed. It will be argued 
that these two boundary conditions do not specify the solution 
uniquely due to a degree of freedom at the horizon (the order of the 
vanishing there can be essentially arbitrary which corresponds to a 
freedom of choosing the radial coordinate). Thus we use the Komar 
integral to ensure that the mass computed via the Komar integral at 
the horizon coincides with the ADM mass. This uniquely specifies the 
solution to the Ernst equation.

\begin{widetext}
In the corotating coordinates introduced in the previous section, 
the determinant of $\mathrm{h}$ does not vanish at the 
horizon. The Ernst equation reads
\begin{equation}
    f\left(\mathcal{E}_{rr}+\frac{1}{(r-m)^{2}-m^{2}
    \cos^{2}\varphi} 
    \left(2(r-m)\mathcal{E}_{r}+\mathcal{E}_{\vartheta\vartheta} +
    \cot\vartheta \mathcal{E}_{\vartheta}\right)\right) =
    \mathcal{E}_{r}^{2}+\frac{\mathcal{E}_{\vartheta}^{2}}{
    (r-m)^{2}-m^{2}
    \cos^{2}\varphi}
    \label{boyer4}.
\end{equation}
Rescaling the coordinates in a way that the radius of the horizon is 
equal to 1, $\tilde{r}= r/(2m\cos^{2}\frac{\varphi}{2})$, we get 
\begin{equation}
    f\left(\mathcal{E}_{\tilde{r}\tilde{r}}+\frac{1}{(\tilde{r}-1)
    (\tilde{r}-\tan^{2}\frac{\varphi}{2})} 
    \left(\left(2\tilde{r}-\frac{1}{\cos^{2}\frac{\varphi}{2}}\right)
    \mathcal{E}_{\tilde{r}}+\mathcal{E}_{\vartheta\vartheta} +
    \cot\vartheta \mathcal{E}_{\vartheta}\right)\right) =
    \mathcal{E}_{\tilde{r}}^{2}+\frac{\mathcal{E}_{\vartheta}^{2}}{
    (\tilde{r}-1)
    (\tilde{r}-\tan^{2}\frac{\varphi}{2})}
    \label{boyer4a}.
\end{equation}
\end{widetext}
Since the real part of the Ernst potential vanishes for $\tilde{r}=1$ 
as $\tilde{r}-1$, the left-hand and the right-hand side of the 
equation are well-behaved at the horizon. Note that the extreme Kerr 
solution for $\varphi=\pi/2$ corresponds to a higher order 
singularity of the equation since $\tan (\varphi/2)=1$ in this case. 
This is the reason why it is numerically challenging to reach the 
extreme Kerr solution in this setting.

The idea is to solve the Ernst equation for the Kerr solution with 
boundary data at the horizon and at some finite outer radius.  The 
problem with this approach is that the horizon is a singular surface of 
the Ernst equation, and that a regularity condition at the horizon does not 
uniquely specify the solution. This can be seen best at the example of the 
Schwarzschild solution, i.e., Kerr for $\varphi=0$. In this case one 
gets for the Ernst equation ($b=0$, no $\vartheta$ dependence)
$$(\ln f)_{rr}+\frac{2r-1}{r(r-1)}(\ln f)_{r}=0.$$ This equation has 
the general solution
\begin{equation}
    f = c_{1}\left(\frac{r-1}{r}\right)^{c_{2}}
    \label{schwarz},
\end{equation}
where $c_{1}$, $c_{2}$ are constants. The constant $c_{1}$ will be 
fixed at infinity or the outer boundary condition, but it can be seen 
that the condition $f=0$ will not fix $c_{2}$ which does not have to 
be integer. Thus the solution is not uniquely specified by the above 
conditions. This is also 
the case for $\varphi\neq0$. The Ernst potential is invariant under 
multiplication by a real constant. This freedom is fixed by the outer 
boundary condition. Looking for a formal solution to the Ernst 
equation in terms of a power series in $r-1$ near the horizon, 
$f=f_{0}(\vartheta)(r-1)^{n_{f}}+\ldots$, 
$b=b_{0}(\vartheta)(r-1)^{n_{b}}+\ldots$, we find for all $n_{b}>n_{f}$
\begin{equation}
    n_{b}=2n_{f}, 
    \label{nbf}
\end{equation}
whereas $n_{f}\in \mathbb{R}$ with $n_{f}>1/2$ and $f_{0}$, $b_{0}$ 
are free.  This corresponds to a freedom in the choice of the radial 
coordinate, $\tilde{r}-1\mapsto (\tilde{r}-1)^{c}$ with $c$ an arbitrary positive 
constant. Thus one has to formulate the boundary value problem in a 
way that $n_{f}=1$ is enforced in order to get the Kerr solution in 
the wanted form. 

To a certain extent, the above non-uniqueness at the horizon is 
addressed by 
using dependent variables of the form $f = (\tilde{r}-1)F$ and 
$b=(\tilde{r}-1)^{2}B$. If $F$ and $B$ are finite at the horizon, the 
minimal order of the vanishing of the Ernst potential there is at 
least assured. With $x=\cos \vartheta$, 
the Ernst equation takes in this case the form
\begin{widetext}
\begin{eqnarray}
    F\left\{(\tilde{r}-1)(\tilde{r}-\tan^{2}(\varphi/2))F_{\tilde{r}\tilde{r}} + (2\tilde{r}-1-\tan^{2}(\varphi/2))F_{\tilde{r}} + F + (1-x^{2})F_{xx} -2x F_{x}\right\}- &  & 
    \nonumber\\
    (\tilde{r}-1)(\tilde{r}-\tan^{2}(\varphi/2))(F_{\tilde{r}}^2 - 
    ((\tilde{r}-1)B_{\tilde{r}}+2B)^2)-(1-x^{2)}(F_{x}^2-(\tilde{r}-1)^2B_{x}^2) & = & 0
    \label{ernstF},
\end{eqnarray}
and
\begin{eqnarray}
    F\left\{(\tilde{r}-1)(\tilde{r}-\tan^{2}(\varphi/2))B_{\tilde{r}\tilde{r}} + (4\tilde{r}-1-3\tan^{2}(\varphi/2))B_{\tilde{r}} + 2B + (1-x^{2})B_{xx} -2x B_{x}\right\}- &  & 
    \nonumber\\
    2(\tilde{r}-\tan^{2}(\varphi/2))F_{\tilde{r}}  
    ((\tilde{r}-1)B_{\tilde{r}}+2B)-2(1-x^{2})F_{x}B_{x} & = & 0
    \label{ernstB}.
\end{eqnarray}
\end{widetext}
This is the form of the Ernst equation to be solved in the following 
sections numerically.

It turns out that this form of the equations still does not have the 
Kerr solution as the unique solution. Therefore we consider in the 
following the Komar integral associated to the Killing vector with 
components $\xi^{k}$
\begin{equation}
    \int_{0}^{2\pi}\int_{0}^{\pi}g_{ik,1}g^{i0}g^{11}\sqrt{-g}d\vartheta d\varphi
    \label{komar0}.
\end{equation}
With 
\begin{equation}
    g^{11}\sqrt{-g}=(\tilde{r}-1)(\tilde{r}-\tan^{2}(\varphi/2))\sin\vartheta = \frac{h_{33}}{\sin\vartheta}
    \label{komar1}
\end{equation}
and 
\begin{equation}
    g^{00}=-\frac{1}{f}+\frac{fa^{2}}{h_{33}},\quad 
    g^{03}=-\frac{af}{h_{33}}
    \label{komar2},
\end{equation}
(this relation also holds in the rotating coordinate system)
one gets for the Killing vector $\partial_{t}$
\begin{eqnarray}
    g_{ik,1}g^{ik}&=&g_{03,r}g^{03}+g_{00,r}g^{00}=(\ln 
    f)_{r}+\frac{aa_{r}f^{2}}{h_{33}}\nonumber \\
    &=&(\ln 
    f)_{r}-\frac{ab_{\vartheta}\sin\vartheta}{h_{33}}
    \label{komar3}.
\end{eqnarray}
At the horizon one has
\begin{align}
    f & =
    -\frac{\tan^{2}(\varphi/2)\sin^{2}\vartheta}{1+\tan^{2}(\varphi/2)\cos^{2}\vartheta}
    \nonumber\\
    a & = -2m\cot(\varphi/2)
    \nonumber\\
    f_{r} & 
    =\frac{1}{2m\cos^{4}(\varphi/2)}\frac{1-\tan^{2}(\varphi/2)\cos^{2}\vartheta}{
    (1+\tan^{2}(\varphi/2)\cos^{2}\vartheta)^{2}}
    \nonumber\\
    a_{r} & 
    =-\frac{1-\tan^{2}(\varphi/2)\cos^{2}\vartheta}{\sin^{2}(\varphi/2)
    \tan(\varphi/2)\sin^{2}\vartheta}.
    \label{hor}
\end{align}
Thus one gets for the Komar mass the well known value
$$\frac{4\pi 
m}{\cos^{2}(\varphi/2)}\int_{0}^{\pi}\frac{1-\tan^{2}(\varphi/2)\cos^{2}\vartheta}{(1+\tan^{2}(\varphi/2)\cos^{2}\vartheta)^{2}}
\sin\vartheta d\vartheta=8\pi m.$$

There is a second Killing vector in the Kerr metric, 
$\partial_{\phi}$, for which the Komar integrand reads
$$(g_{03,r}g^{00}+g_{33,r}g^{03})\sqrt{-g}g^{11},$$
which at the horizon takes the form
\begin{widetext}
$$\frac{a}{\sin\vartheta}(a_{r}af^{2}-h_{33,r})=4m^{2}\cot(\varphi/2)\sin\vartheta
\left(\frac{1}{\cos^{2}(\varphi/2)}\frac{1-\tan^{2}(\varphi/2)\cos^{2}\vartheta}{(1+\tan^{2}(\varphi/2)\cos^{2}\vartheta)^{2}} +1-2\cos^{2}(\varphi/2)\right).$$
\end{widetext}
Integrating we find $16\pi m^{2}\sin\varphi=8\pi J$.

The Komar mass can be also computed in corotating coordinates. 
However there one has that $a'f'$ vanishes as 
$f'$ at the horizon, see (\ref{afp}). In this case the integrand of 
the Komar mass reads at the horizon 
$-(1-\tan^{2}(\varphi/2))\sin\vartheta$, i.e., it does not contain 
information on the function $F$ at the horizon. We get 
$$4\pi m \cos^{2}(\varphi/2)(1-\tan^{2}(\varphi/2))\int_{-1}^{1}dx = 
8\pi(m+\Omega J).$$ 
Thus it is not useful to assure a non-vanishing of $F$ at the horizon 
by imposing the value $8\pi(m+\Omega J)$ for the Komar integral in numerical 
computations. But it will allow in the case of binary black holes 
with a helical Killing vector to relate the values computed at the 
horizon to (asymptotically defined) multipoles of an asymptotically 
flat spacetime imposed at the outer computational boundary.

The Komar integral for the 
Killing vector $\partial_{\phi}$ reads 
\begin{equation}
    2\pi\int_{0}^{\pi}\left(a_{r}(h_{33}+a^{2}f^{2})+h_{33}a(2(\ln 
f)_{r}-(\ln h_{33})_{r})\right)\frac{d\vartheta}{\sin\vartheta}.
    \label{komarphip}
\end{equation}
On the horizon the integrand reduces to 
$2m\cos^{2}(\varphi/2)a(R,\vartheta)(1-\tan^{2}(\varphi/2))\sin\vartheta$. Since at the 
horizon 
\begin{eqnarray*}
    a' & = & 
    \frac{2m}{\sin(\varphi/2)\cos^{3}(\varphi/2)(1-\tan^{2}(\varphi/2))}\times\\
     &  & \frac{1-\tan^{2}(\varphi/2)\cos^{2}\vartheta}{(1+\tan^{2}(\varphi/2)\cos^{2}\vartheta)^{2}}
-2m\cot(\varphi/2),
\end{eqnarray*}
we get for the Komar integral as before $8\pi J$. This condition will 
be imposed in the numerical solution of the Ernst equation.

To compute this integral, the function $a$ has to be known, and this 
implies that a constant in $b$ is fixed on the horizon. The function 
$a$ can be computed from (\ref{btheta}),
\begin{equation}
    a_{\tilde{r}}= -\frac{B_{\vartheta}}{F^{2}}2m\cos^{2}(\varphi/2)\sin\vartheta
    \label{komar4}
\end{equation}
and from (\ref{br})
\begin{equation}
    a_{\vartheta}=-\frac{(\tilde{r}-1)B_{\tilde{r}}+2B}{F^{2}}2m\cos^{2}(\varphi/2)(\tilde{r}-\tan^{2}(\varphi/2)
    \sin\vartheta
    \label{komar5}.
\end{equation}

\section{Numerical approaches to the Ernst equation}
In this section we outline the numerical approaches to solve 
the Ernst equation in the form  (\ref{ernstF}) and (\ref{ernstB}). To 
approximate derivatives, we use a pseudospectral approach in 
$\tilde{r}$ and $x$ based on discretizing both coordinates. The 
resulting system of finite dimension for the discretized $F$ and $B$ is then solved 
with a Newton-Armijo iteration. 

\subsection{Polynomial interpolation and differentiation matrices}

To solve the Ernst equation, we need to approximate numerically the derivative of a function $\mathcal{F}: 
[-1,1]\mapsto \mathbb{C}$. To this end we use polynomial 
interpolation as detailed 
for instance in \cite{trefethen}. We 
introduce  on $[-1,1]$,  the $N+1$ \emph{Chebyshev collocation points} 
\begin{equation}
    l_{j}=\cos\left(\frac{j\pi}{N}\right),\quad j=0,\ldots,N
    \label{colpoints},
\end{equation}
where $N$ is some natural number.  The
Lagrange polynomial $p(l)$ of order $N$ satisfying the relations
$p(l_{j})=\mathcal{F}(l_{j})$, $j=0,\ldots,N$ is then constructed. The derivative of 
$\mathcal{F}$ at 
the collocation points $l_{j}$ is approximated via the derivative of this 
polynomial, 
$$\mathcal{F}'(l_{j})\approx p'(l_{j})=:\sum_{k=0}^{N}D_{jk}\mathcal{F}(l_{k}),$$
where $D$ is a \emph{differentiation matrix}. The 
matrices $D$ for Chebyshev collocation points are given in  
\cite{trefethen}, a Matlab code to generate them can be found at 
\cite{trefethenweb}. Second derivatives of the function $\mathcal{F}$ will 
be approximated by $D^{2}\mathrm{F}$, where 
$\mathrm{F}_{j}=\mathcal{F}(l_{j})$, $j=0,\ldots,N$. 
This method is known to show \emph{spectral convergence} for analytic 
functions, i.e., an exponential decrease of the numerical error with 
$N$. 

This
 pseudo-spectral approach is equivalent to an approximation 
of the function $\mathcal{F}$ by a (truncated) series of Chebyshev polynomials 
$T_{n}(l)$, $n=0,\ldots,N$, where
\begin{equation}
    T_{n}(l)=\cos(n\mbox{arccos}(l))
    \label{cheb}.
\end{equation}
A \emph{Chebyshev collocation method} consists in approximating 
$\mathcal{F}$ via $ 
\sum_{n=0}^{N}c_{n}T_{n}(l)$, where the 
\emph{spectral coefficients} $c_{n}$ are given by,
\begin{equation}
    \mathcal{F}(l_{j})=\sum_{n=0}^{N}c_{n}T_{n}(l_{j}),\quad j=0,\ldots,N
    \label{coll}.
\end{equation}

Note that because of (\ref{cheb}), the coefficients $c_{n}$ in (\ref{coll}) 
can be computed via a 
\emph{fast cosine transformation} (fct) which is closely related to 
the \emph{fast Fourier transform} (fft), see 
\cite{trefethen}. Since the fct is in 
contrast to the fft not a precompiled command in Matlab being used 
here, it is considerably slower than 
the latter. Thus we apply here the pseudospectral approach in 
computations. But the fct allows to control the resolution in terms 
of the Chebyshev coefficients: as for Fourier coefficients of real 
analytic functions, it is known that Chebyshev coefficients of such 
functions decrease exponentially with $n$. This allows to control 
that the computed functions have the expected analyticity properties. 
In addition it permits to control the resolution of the solution in 
terms of Chebyshev polynomials: if the Chebyshev coefficients 
decrease to machine precision (here $10^{-16}$, in practice limited 
to roughly $10^{-14}$ because of unavoidable rounding errors), 
maximal resolution with this approach has been reached. 

\begin{figure}[htb!]
   \includegraphics[width=0.49\textwidth]{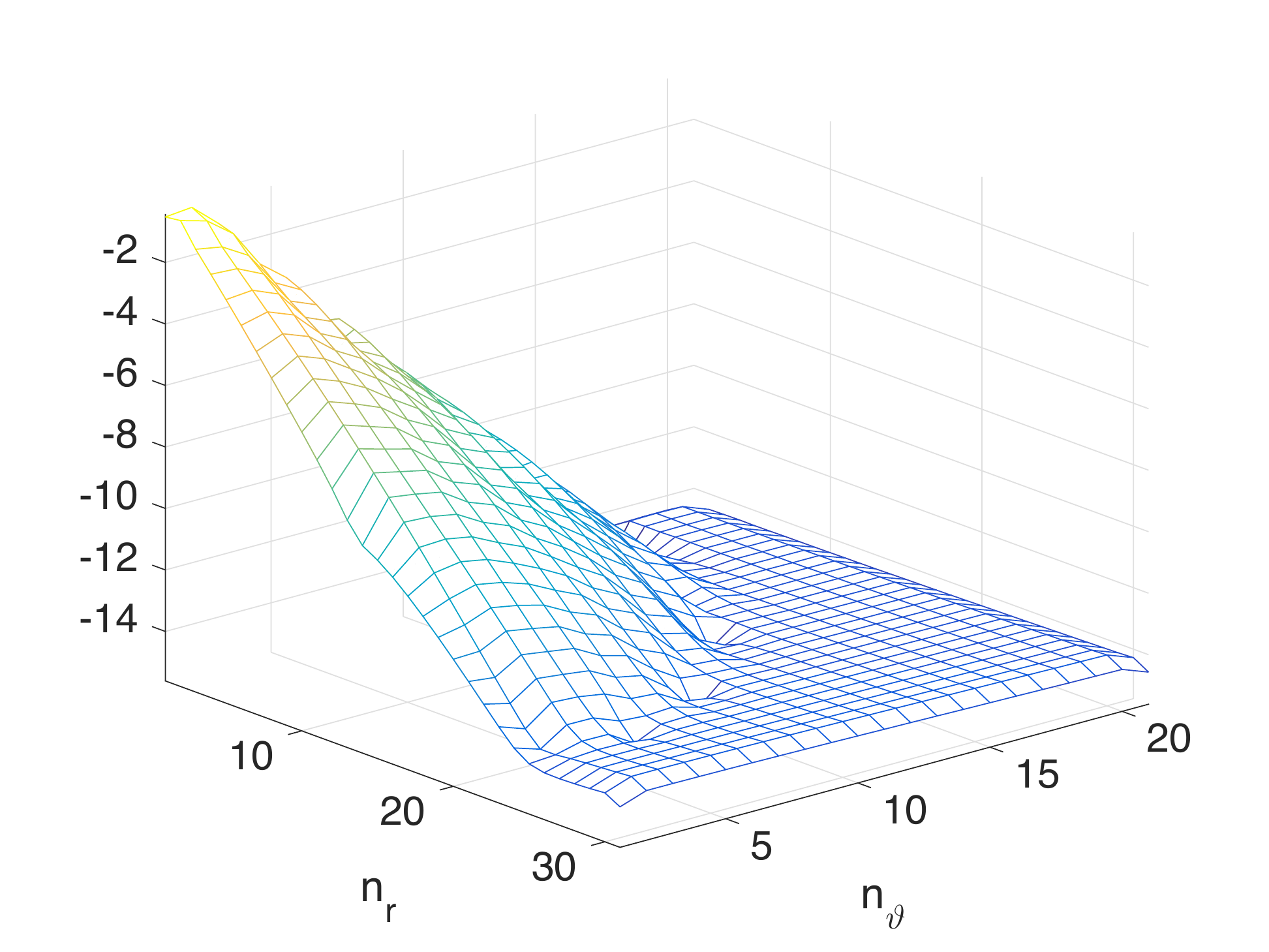}
 \caption{Logarithm of the Chebyshev coefficients of the Ernst potential of the Kerr 
 solution in rotating coordinates for $\varphi=1$. }
 \label{Echeb}
\end{figure}

For the Ernst equation, we discretize $r\in[1,R]$, where $R$ 
is the radius of the outer boundary, via  
$r_{j}=R(1+l_{j})/2+(1-l_{j})/2$, $j=0,1,\ldots,N_{r}$ with the 
$l_{j}$ from (\ref{colpoints}). Similarly we discretize the 
coordinate $x=\cos \vartheta$. Since the Kerr solution is 
axisymmetric, it is sufficient to consider $\theta\in[0,\pi/2]$. Thus 
we can write $x_{j}=(1+l_{j})/2$ $j=0,1,\ldots,N_{\vartheta}$ with 
the $l_{j}$ from (\ref{colpoints}). For the Ernst potential of the Kerr solution (\ref{fp}) 
and (\ref{bp}) with $\varphi=1$ we get the Chebyshev coefficients 
shown in Fig.~\ref{Echeb}. It can be seen that the coefficients 
decrease to machine precision with $N_{r}=30$ and $N_{\vartheta}=20$. 

\subsection{Newton-Armijo iteration}
The discretisation introduced in the previous section gives matrices 
with components
$F(\tilde{r}_{j},x_{k}$), $B(\tilde{r}_{j},x_{k})$, $j=0,\ldots,N_{r}$, 
$k=0,\ldots,N_{\vartheta}$. These are combined to a vector 
$\mathrm{G}$ of length $2(N_{r}+1)(N_{\vartheta}+1)$. This 
discretisation implies that the equations (\ref{fp}) and (\ref{bp}) 
are discretized in the same way. The discretized 
equations can be combined to a system of 
$2(N_{r}+1)(N_{\vartheta}+1)$ nonlinear equations of the form 
$\mathcal{G}(\mathrm{G})=0$.

This system of equations will be solved iteratively with a 
Newton-Armijo method giving iterate $n+1$ in dependence of iterate 
$n$, 
\begin{equation}
    \mathrm{G}_{n+1}=\mathrm{G}_{n}-\lambda\mbox{Jac}^{-1}\mathcal{G}(\mathrm{G}_{n})
    \label{NA},
\end{equation}
where $\mbox{Jac}$ is the Jacobian of $\mathcal{G}$ with respect to 
$\mathrm{G}$ taken for $\mathrm{G}=\mathrm{G}_{n}$, and where $0< 
\lambda\leq 1$ is a parameter  (to be discussed below) equal to 1 in the standard Newton 
iteration. Note that the Jacobian is a matrix of order 
$2(N_{r}+1)(N_{\vartheta}+1)\times 2(N_{r}+1)(N_{\vartheta}+1)$. It 
is known that the standard Newton iteration ($\lambda=1$) converges 
quadratically. But the convergence is local, i.e., the initial 
iterate $\mathrm{G}_{n}$ has to be close to the exact solution to 
ensure convergence,  see for instance the discussion in 
\cite{newtonit} and references therein. 

If the initial iterate $\mathrm{G}_{0}$ is not close enough to the 
solution, the standard Newton iteration fails in general to converge.
A simple approach is to use some \emph{relaxation} in the 
iteration, i.e., choose a value of $\lambda<1$. This can restore 
convergence, but the quadratic convergence will be lost. Whereas this 
could be acceptable in the two-dimensional setting studied here, it 
certainly will not be in the three-dimensional problem for which this 
is a test case. But as will be discussed in the following section, 
even here one might be forced to use prohibitively small values of 
$\lambda$ to avoid divergence. Therefore we apply  an Armijo 
approach, i.e., a dynamical adjustment of the value of $\lambda$.

For each $\mathrm{G}_{n}$ in the iteration, the norm 
$\mathcal{N}_{n}:=||\mathcal{G}(\mathrm{G}_{n})||_{\infty}$ is computed. If 
$\mathcal{N}_{n+1}<\epsilon$, where $\epsilon$ 
is the prescribed aimed at accuracy (in our case typically $10^{-10}$ 
or smaller) the iteration is stopped. If this is not the case, it is 
checked whether
$\mathcal{N}_{n+1}<(1-\alpha \lambda)
\mathcal{N}_{n}$. Here $\alpha$ is some 
constant which is chosen to be $10^{-4}$. If the condition is 
met,  the new iterate 
$\mathrm{G}_{n+1}$ is computed via (\ref{NA}) without changing the 
current value of $\lambda$.   

If this is not the case, i.e., if the new iterate would give a worse 
(up to a factor $\alpha\lambda$ which can be freely chosen)
solution to the equation $\mathcal{G}(\mathrm{G})=0$ than the 
previous one, a \emph{line search} is performed: first the value of 
$\lambda$ is halved to give a $\tilde{\lambda}$, and the corresponding value of 
$\mathcal{N}_{l}:=||\mathcal{G}(\mathrm{G}_{n+1}^{l})||_{\infty}$ is computed, where 
$\mathrm{G}_{n+1}^{l}$ is the $\mathrm{G}_{n+1}$ of (\ref{NA}) with 
the current value of $\lambda$. If this norm is smaller than $(1-\alpha \lambda)
\mathcal{N}_{n}$, the new value is 
determined by a fitting to a quadratic model: the interpolation 
polynomial passing through the three values of the norms 
$\mathcal{N}_{n}$ for $\lambda=0$, $\mathcal{N}_{n+1}$ for 
$\lambda=\lambda_{o}=1$ and $\mathcal{N}_{l}$ for $\lambda=\tilde{\lambda}$ reads 
$$p(\lambda)=\frac{(\lambda-\lambda_{o})(\lambda-\tilde{\lambda})}{\lambda_{o}\tilde{\lambda}}\mathcal{N}_{n}
+\frac{(\lambda-\lambda_{o})\lambda}{(\tilde{\lambda}-\lambda_{o})\tilde{\lambda}}\mathcal{N}_{l}
+\frac{\lambda(\lambda-\tilde{\lambda})}{\lambda_{o}(\lambda_{o}-\tilde{\lambda})}\mathcal{N}_{n+1}.$$
The minimum value $\lambda_{m}$ of this polynomial is taken as the 
new $\tilde{\lambda}$ unless it is smaller than $\lambda/10$ (in this case 
$\lambda/10$ is taken) or larger than $\lambda/2$ (in this case 
$\lambda/2$ is taken). If the condition $\mathcal{N}_{l}<(1-\alpha 
\tilde{\lambda})\mathcal{N}_{n}$ is still not met, the above approach 
is iterated with the new $\tilde{\lambda}$ and $\lambda_{o}$ replaced 
by the old value of $\tilde{\lambda}$. The line search is stopped if the current 
value of $\tilde{\lambda}$ is smaller than $10^{-2}$.

The next step 
of the Newton 
iteration  is then started again with $\lambda=1$. For details of 
the approach, the reader is referred to \cite{newtonit}. The method 
considerably generalizes the admissible choices for the initial 
iterate to achieve convergence of the Newton iteration. But obviously the 
closer this initial iterate is to the wanted the solution, the more 
rapid will be the convergence. 

\subsection{Boundary values and Komar integral}
At the boundaries of the computational domain, it might be necessary 
to impose boundary conditions in order to avoid a degenerate Jacobian 
in (\ref{NA}). On the axis $\vartheta=\pi/2$ and at the horizon 
$\tilde{r}=1$, this is not necessary 
since the equations (\ref{ernstF}) and (\ref{ernstB}) are singular there. 
Thus the condition of regularity  determines the solution 
there, and instead of a boundary condition, just the PDE can be 
imposed. 

The situation is different at the outer boundary $r=R$ and in the 
equatorial plane $\vartheta=0$ where boundary conditions have to be 
enforced. At the former, just the exact Kerr solution in corotating 
coordinates (\ref{fp}) and (\ref{bp}) will be imposed. Alternatively 
the asymptotic solution (\ref{asym2}) and (\ref{asym2b}) could be 
prescribed as will be done in the binary case. In the equatorial plane, 
we use just the equatorial symmetry of the solution which implies 
$F_{\vartheta}(\tilde{r},0)=b(\tilde{r},0)=0$.

The conditions at the outer boundary and in the equatorial plane will 
be implemented via Lanczos $\tau$-method \cite{tau}. The idea is to 
eliminate parts of the equation 
$\mbox{Jac}(\mathrm{G}_{n+1}-\mathrm{G}_{n})+\mathcal{G}(\mathrm{G}_{n})=0$ 
and to replace them with the boundary conditions. Thus we replace the 
equations corresponding to $\tilde{r}=R$ by the Kerr solution there, 
and the equations corresponding to the equatorial plane by the 
symmetry conditions on the Ernst potential. The derivative with 
respect to $x$ is approximated as all $x$ derivatives with the 
corresponding differentiation matrix. It is known, see e.g. 
\cite{trefethen}, that the $\tau$ method does not implement the 
boundary conditions exactly, but with the same spectral accuracy as 
the solution of the PDE is approximated. 

In a similar way the Komar integral (\ref{komarphip}) is imposed. To 
determine the integrand at the horizon, we numerically integrate 
(\ref{komar5}) to determine the function $a$. This is done by 
inverting the matrix $D_{x}$ (the differentiation matrix 
corresponding to the coordinate $x$) with a vanishing boundary condition 
implemented at the 
horizon via a $\tau$-method. The integral over the horizon is then 
computed with the \emph{Clenshaw-Curtis method}: as already 
mentioned, the polynomial interpolation on Chebyshev collocation 
points is equivalent to a Chebyshev collocation method (\ref{coll}). 
Thus if an integrand is expanded in terms of Chebyshev 
polynomials, $$\int_{-1}^{1}\mathcal{F}(l)dl\approx 
\sum_{n=0}^{N}c_{n}\int_{-1}^{1}T_{n}(l)dl=\sum_{n=0}^{N}
w_{n}\mathcal{F}(l_{n})$$ (the last 
step following from the collocation method (\ref{coll}) relating 
$c_{n}$ and $\mathcal{F}(l_{n})$) where the 
$w_{n}$, $n=0,\ldots,N$ are some known weights (see 
\cite{trefethenweb} for a Matlab code to generate them).
The Clenshaw-Curtis scheme is also a spectral method.

The condition that the Komar integral is equal to $8\pi J$ is again 
imposed via a $\tau$-method as above. But this time an equation has 
to be replaced in (\ref{NA}) which is not redundant as before. We 
generally take the equation corresponding to (\ref{ernstB}) on the 
intersection of the horizon and the axis or the intersection of 
horizon and equatorial place. Thus the Komar integral 
will be implemented in the iteration in the same way as the boundary 
conditions. 

\section{Examples}
In this section, the numerical approach detailed in the previous 
section will be tested for various initial iterates 
for various values of the parameter $\varphi$. Generally the 
iteration converges more rapidly the smaller $\varphi$ is, i.e., the 
farther the solution is from the extreme Kerr solution.

\begin{figure}[htb!]
   \includegraphics[width=0.49\textwidth]{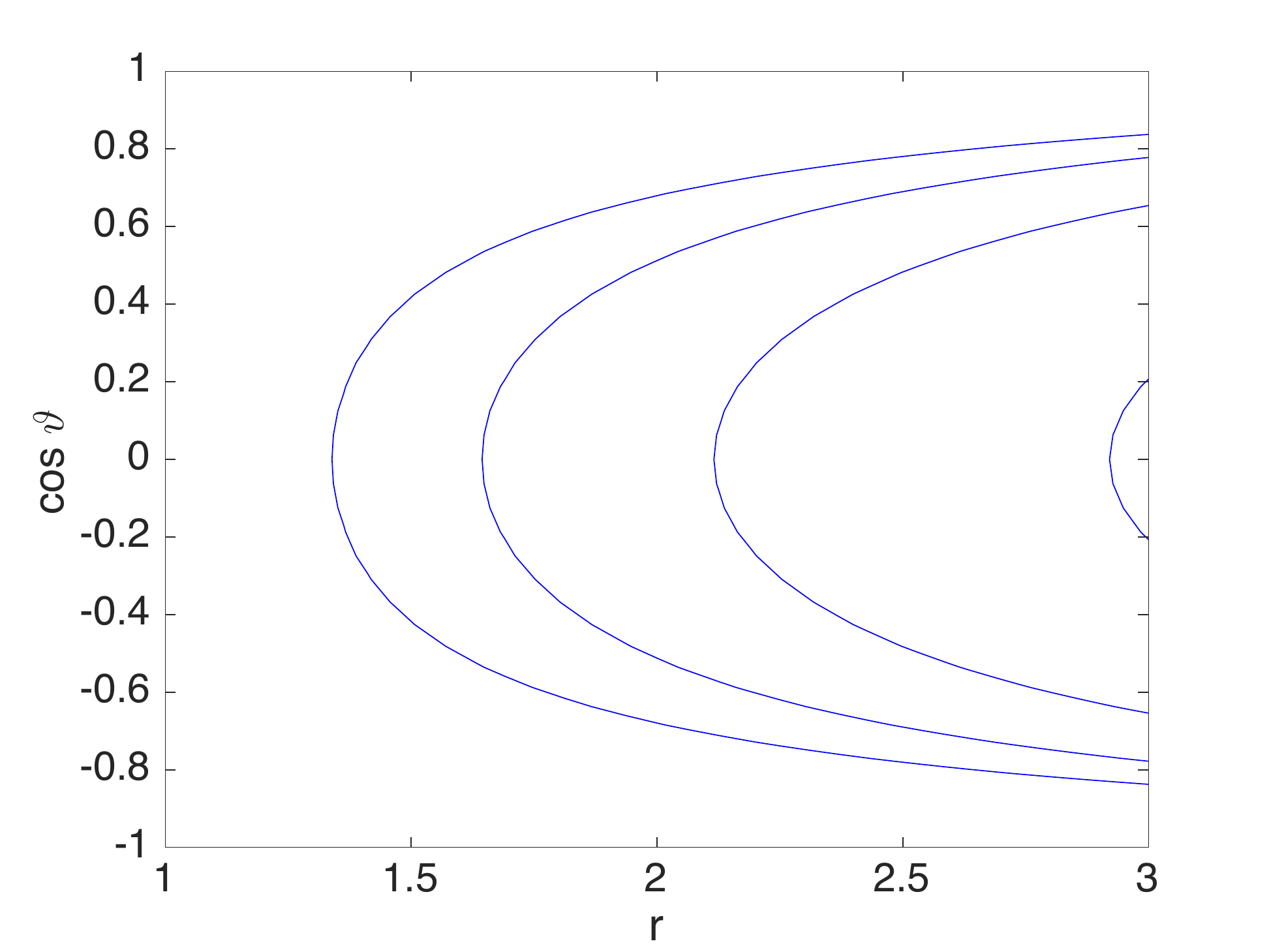}
 \caption{Light cylinders of the Kerr solution (\ref{fp}) in 
 corotating coordinates for the values of $\varphi=0.6,0.8,1.0,1.2$ 
 from right to left. }
 \label{lightcylinder}
\end{figure}

In this section we always choose the outer radius $R=3$. In Fig. 
\ref{lightcylinder}, it can be seen that the light cylinder will be 
for values of $\varphi$ larger than $0.5$ in the computational zone. 
Convergence of the scheme in this case will indicate that the light 
cylinder does not pose an insurmountable problem for the iteration. 

Throughout this section we work with 
$N_{r}=31$ and $N_{\vartheta}=20$ collocation points, numbers which 
ensure the necessary resolution as shown by  Fig.~\ref{Echeb}: 
the Chebyshev coefficients decrease to 
machine precision for $\varphi\leq 1$.  
Since we impose the Komar 
integral in order to obtain a unique solution at the horizon as 
discussed before, the condition for the Komar integral replaces one 
of the equations in (\ref{NA}). This procedure eventually leads 
generically to a 
unique solution (for special values of the parameters, there can be 
still other solutions to the Ernst equation satisfying the boundary 
conditions), but it can destabilize the iteration if the initial 
iterate is too far from the wanted solution.  This 
is also the reason why we do not present a study of the dependence on 
the parameters $N_{r}$ and $N_{\vartheta}$. For smaller values 
of these, the iteration will not converge because of the imposed 
Komar integral replacing one of the equations. And without this 
integral, the solution to the Ernst equation will not be unique. 

We first consider several initial iterates for the case $\varphi=1$ 
for which according to Fig. \ref{lightcylinder} the light cylinder 
extends through most of the computational domain. 
This is already a fast spinning black 
hole and thus provides a good test of the scheme. As the first initial 
iterate we take a factor $\lambda$ times the exact Kerr solution. It can be 
seen in Table \ref{tab:phi1} that for both $\lambda=0.9$ and 
$\lambda=1.1$ the iteration converges without any line search to the 
order $10^{-12}$ (the code is stopped as soon as the residual drops 
below $10^{-11}$). The $L^{\infty}$ norm of the difference between 
numerical and exact 
solution is in this case of the order of $10^{-12}$.  A similar 
behavior is observed if the exact Kerr
solution plus $0.1$ times a Gaussian in $\tilde{r}$ is taken as initial 
iterate.
If we take as the 
initial iterate the exact Kerr solution for $\varphi=0.9$, the 
iteration converges after 5 iterations to the order of $10^{-12}$. 
This appears to be the optimal accuracy reachable with the 
approach. The above examples show that the iteration is stable and converges 
rapidly for various initial iterates to an accuracy of better than 
$10^{-11}$ both as a residual to the numerically implemented 
equations and compared to the exact solution. 
\begin{table}
\begin{tabular}{|c|c|c|}
    \hline
    initial iterate & iterations & $||u-u_{Kerr}||_{\infty}$  \\
    \hline
    $u_{Kerr}(\varphi=0.9)$ & 7 & $1.9*10^{-12}$  \\
    \hline
    $1.1u_{Kerr}(\varphi=1)$ & 5 & $1.7*10^{-12}$  \\
    \hline
   $u_{Kerr}(\varphi=1)+0.1\exp(-\tilde{r}^{2})$ & 6 & $1.7*10^{-12}$  \\
    \hline
   $0.9u_{Kerr}(\varphi=1)$ & 5 & $2.1*10^{-12}$  \\
    \hline
\end{tabular}
\caption{\label{tab:phi1}A table of the convergence of the iteration 
scheme for $\varphi=1$ and various initial iterates.}

\end{table}

The used form of the Ernst equation does not allow to treat the 
extreme Kerr solution in this way. The reason for this is that the 
horizon no longer corresponds to a regular singularity of the 
equations in this case, 
and that the light cylinder touches the horizon. 
It would be necessary to address this case explicitly, but this is not 
the goal here. However, it is interesting to note that one gets rather 
close to the extreme Kerr solution. One has just to find better 
initial iterates in this case to get convergence. For $\varphi=1.5$, 
we get for an initial iterate of a factor $\lambda$ times the exact 
solution again rapid convergence, see Table \ref{tab:phi1.5}. Note that in this case the exact 
location of the light cylinder is the correct one of the wanted Kerr 
solution. The situation is similar for an  initial iterate of the 
exact solution plus a small Gaussian in $\tilde{r}$. But only for a 
Gaussian of maximum $0.01$, not for $0.1$ as before in Table 
\ref{tab:phi1}. There is also no convergence if the exact Kerr 
solution with $\varphi=1.4$ is taken, one has to be as close as 
$\varphi=1.495$. In all cases the  $L^{\infty}$ 
norm of the difference between numerical and exact solution is of the 
order of  $10^{-10}$. 
\begin{table}
\begin{tabular}{|c|c|c|}
    \hline
    initial iterate & iterations & $||u-u_{Kerr}||_{\infty}$  \\
    \hline
   $0.9u_{Kerr}(\varphi=1.5)$ & 14 & $2.0*10^{-10}$  \\
    \hline
    $1.1u_{Kerr}(\varphi=1.5)$ & 8 & $2.1*10^{-10}$  \\
    \hline
    $u_{Kerr}(\varphi=1.495)$ & 10 & $2.1*10^{-10}$  \\
    \hline
   $u_{Kerr}(\varphi=1.5)+0.01\exp(-\tilde{r}^{2})$ & 6 & $2.0*10^{-10}$  \\
    \hline
\end{tabular}
\caption{\label{tab:phi1.5}A table of the convergence of the iteration 
scheme for $\varphi=1.5$ and various initial iterates.}
\end{table}

The above results indicate that it might be possible to start the 
iteration close to the static Schwarzschild solution and use the
found numerical solution for a given value of $\varphi$ as initial 
iterates for larger values of $\varphi$. This would allow to increase 
the angular momentum of the 
black hole in the iterations. The steps have to be smaller the closer one is to the 
extreme black hole. Problems in this approach are obviously related 
to the location of the light cylinder. Whereas  an initial iterate of 
the form of the exact Kerr solution multiplied by some factor leads 
to rapid convergence, this is not the case for an initial iterate 
with a clearly different form of the light cylinder. This is not 
surprising since the latter corresponds to a singularity of the 
Ernst equation. In fact in can be seen in Table \ref{tab:phi} that 
the iteration converges rapidly up to values of $\varphi=0.5$ if the 
initial iterate is the Kerr solution with $\varphi-0.1$. It is clear 
from Fig.~\ref{lightcylinder} that in these cases, there is no light 
cylinder in the computational domain. Convergence problems appear for 
$\varphi=0.6$ and $\varphi=0.7$ for which the light cylinder appears 
very close to the outer computational boundary $R=3$. Here the 
initial iterate must be close to the final solution, i.e., the light 
cylinder should be close to its exact location. For larger values 
of $\varphi$, the light cylinder is not only localized close to the 
boundary as can be seen in Fig.~\ref{lightcylinder}, and the 
iteration converges again rapidly. Thus it appears that special care has to 
be taken in the choice of the initial iterate if the light cylinder 
appears only close to the outer boundary of the computational domain, 
or if it is close to the horizon as in almost extreme black holes. 
\begin{table}
\begin{tabular}{|c|c|c|}
    \hline
    $\varphi$ & iterations & $||u-u_{Kerr}||_{\infty}$  \\
    \hline
    0.1 & 6 & $4.8*10^{-12}$  \\
    \hline
    0.2 & 9 & $3.3*10^{-12}$  \\
    \hline
    0.3 & 8 & $3.8*10^{-12}$  \\
    \hline
    0.4 & 30 & $2.1*10^{-12}$  \\
    \hline
    0.5 & 9 & $2.5*10^{-11}$  \\
    \hline
    $0.6^{*}$ & 24 & $2.4*10^{-12}$  \\
    \hline
    $0.7^{*}$ & 9 & $2.5*10^{-12}$  \\
    \hline
    0.8 & 6 & $1.3*10^{-11}$  \\
    \hline
    0.9 & 6 & $1.3*10^{-12}$  \\
    \hline
    1 & 7 & $1.9*10^{-12}$  \\
    \hline
\end{tabular}
\caption{\label{tab:phi}A table with the convergence of the iterative 
solution of the Ernst equation for various values of $\varphi$. The 
initial iterate is always the exact Kerr solution for $\varphi-0.1$ 
except for the cases marked with a star: for $\varphi=0.6$, the 
iteration is started with the value $0.51$, for $\varphi=0.7$ with 
$\varphi=0.66$. }
\end{table}

\section{Outlook}
In the previous section we have shown that the Ernst equation in the 
form (\ref{ernstF}) and (\ref{ernstB}) can be solved iteratively for a 
Kerr black hole in a frame corotating with the horizon. The 
numerically challenging part is the location of the light 
cylinder which is a singularity for the equation. The main problems 
arise if the light cylinder is located close to the boundary of the 
computational domain or close to the horizon as in almost extreme 
black holes. It was shown that these difficulties could be addressed 
by performing line searches in the iteration. 

A technical problem of the Ernst equation is the fact that the latter 
is homogeneous in the Ernst potential which implies that with 
$\mathcal{E}$ also a constant times $ \mathcal{E}$ is a solution. In 
addition the order of the vanishing of the Ernst potential at the 
horizon is not uniquely fixed. Thus the 
asymptotic behavior of the Ernst potential together with a regularity 
condition at the horizon does not uniquely identify the solution. 
Therefore we used the Komar integral for the Killing vector 
$\partial_{\phi}$ to establish a unique solution. 

The above results indicate that the Ernst approach should also allow 
a numerical solution in the case of binary black holes with a helical 
Killing vector. The idea is to use adapted coordinates as bispherical 
coordinates in which the horizons are given as constant coordinate 
surfaces, for instance the approach \cite{kadath}. 
The Einstein equations in the projection formalism will be 
solved with a spectral method and an Newton-Armijo approach as in the 
present paper on a finite computational domain. At the boundary of 
the computational domain, a solution to the linearized Einstein 
equations as in \cite{kleinprd} or an asymptotically flat solution 
will be imposed as boundary conditions. This will be the subject of 
further work.

\begin{acknowledgments}
This work has been supported by the ANR via the program 
ANR-09-BLAN-0117-01 and the region of Burgundy. We thank R. Beig,  J. 
Frauendiener, and P. Grandcl\'ement for helpful remarks and hints.
\end{acknowledgments}

\end{document}